\documentclass{ws-p9-75x6-50}
\usepackage{amsfonts,amssymb,epsfig}
\usepackage{amsmath}
\def\sqr#1#2{{\vcenter{\hrule height.#2pt\hbox{\vrule width.#2pt
height#1pt \kern#1pt \vrule width.#2pt}\hrule height.#2pt}}}
\def\square{\mathchoice\sqr64\sqr64\sqr{4.2}3\sqr{3.0}3 \ \!} 

\def\br{\begin{eqnarray}}
\def\er{\end{eqnarray}}
\def\brn{\begin{eqnarray*}}
\def\ern{\end{eqnarray*}}
\def\er{\end{eqnarray}}
\def\beq{\begin{equation}}
\def\eeq{\end{equation}}

\def\vp{\varphi}

\def\E{\mathcal{E}}

\def\tt{\tilde {t}}

\def\bpsi{\boldsymbol{\psi}}

\def\dbpsi{\boldsymbol{\dot{\psi}}}
\def\ddbpsi{\boldsymbol{\ddot{\psi}}}

\title{\ On the derivation of the equation of motion in a scalar model.}
\author{Shmuel Kaniel 
 and Yakov Itin }
\address{Institute of Mathematics,  Hebrew University of Jerusalem, \\
Givat Ram, Jerusalem 91904, Israel \\E-mail: kaniel@math.huji.ac.il, itin@math.huji.ac.il}
\begin{document}  
\date{}
\maketitle
General Relativity (GR) is unique among the class of field theories 
in the treatment of the equations of motion. 
The equations of motion of massive particles are completely 
determined by the field equation. \\
 It turns out that most field equations are  
linear in the second order derivatives and quadratic in 
the first order derivatives both with  coefficients that depend
only on the field variables. 
For the purpose  of exposition let us write the general field equation, 
symbolically, as 
\begin{equation}\label{fe2} 
aA(\Phi)+bB(\nabla \Phi, \nabla \Phi)=0,
\end{equation}
where $A$ is a second order linear operator, while $B$ is a quadratic form. 
The coefficients $a$ and $b$ are the dimensionless functions 
of the field variable (or constants)
$a=a(\Phi), \ b=b(\Phi).$\\
The covariance condition as well as the existence of an 
action functional provide severe restrictions on the 
coefficient  $a(\Phi)$ and $b(\Phi)$. 
Note that Einstein equation does have the form (\ref{fe2}).\\ 
A  novel algorithm for the derivation 
of equations of motion is worked. It is for field equations that are 
Lorentz invariant. A condense summary of the algorithm follows.
\begin{itemize}
\item[\bf{1}] Compute a static, spherically symmetric 
solution $F$ of the field equation. 
It will be singular at the origin. 
This will be taken to be the field generated by a single particle.
\item[\bf{2}] Move the solution on a trajectory $ \bpsi(t)$ 
by apply the instantaneous Lorentz transformation based on $\dbpsi(t)$.
\item[\bf{3}] Take the field generated by $n$ particles 
to be the superposition of the fields generated by the  single particles. 
\item[\bf{4}] Compute the leading (linear ) part of the equation. 
Hopefully, only terms that involves $ \ddbpsi$ will be dominant.
It turns out that these terms are {\it the agent of inertia}. 
\item[\bf{5}] Compute the ``force'' between the particles by 
the quadratic part of the equation. 
By {\bf{3}} it will be 
$$bB(\nabla \sum {}^{(j)}F,\nabla \sum {}^{(k)}F).$$ 
Since 
$$aA({}^{(j)}F)+bB(\nabla  {}^{(j)}F,\nabla  {}^{(j)}F)=0$$ 
it follows that 
$$bB(\nabla  {}^{(j)}F,\nabla  {}^{(k)}F) \qquad \textrm {with}  \qquad j\ne k$$
 remains. 
\item[\bf{6}] Equate for each singularity, the highest 
order terms of the singularities that came from the linear 
part and the quadratic parts, respectively. This is an 
equation between the inertial part and the force. 
For the $j$-th singularity the linear terms will contribute $\ddbpsi_j$, 
while the quadratic part will include all the terms with fixed $j$. 
Equating of the two highest order singularities should, hopefully, 
result in Newton's law of attraction.
\end{itemize}
The proposed method  
was applied to the vacuum Einstein field equation. 
$$R_{ik}=0.$$
In the first order approximation the metric is determined by a unique scalar function $f$. The field equation is transformed 
to a nonlinear scalar field equation 
\begin{equation} 
d*d \ \!\! f=kd \ \!\! f\wedge *d \ \!\! f,
\end{equation}
or, in coordinate form
\begin{equation} \label{s-eq}
\square f= k\eta^{ab}f_{,a}f_{b}.
\end{equation}
The algorithm above  resulted, indeed, in the verification of 
Newton gravitation law. \\
The trajectories obtained above are approximate. 
At this point, two avenues are open. 
The first one, which is adopted by EIH is to get higher order approximations 
to the trajectories. 
This procedure is also used in the PPN approach.
By these methods, the successive approximations become highly 
singular near the particle trajectories. \\
The second avenue is to embed the singularities in a field satisfying the 
field equations. 
For that purpose, the successive approximations should add, near 
the trajectories, regular terms and, possibly,  low order 
singular terms as well. 
In the first order approximation the trajectories are taken 
to be fixed and the calculation is done only 
for the first order terms. 
Only the scalar model will be considered.\\
The desired field will be a solution $\vp$ of (\ref{s-eq}),
\begin{equation}\label{4-1}
\vp=e^{-kf+g},
\end{equation}
where $f$ is defined by the singular solution  and $g$ is regular 
at all the singularities (``multi Green function''). 
Thus 
\begin{equation}\label{4-2}
\square g-2k\eta^{ab}f_{,a}g_{,b}+\eta^{ab}g_{,a}g_{,b}=-F,
\end{equation}
where
\begin{equation}\label{4-3}
k(\square f-k\eta^{ab}f_{,a}f_{,b})=F
\end{equation}
Let us do some formal reasoning. The highest order singular terms are, near 
the $j$-th trajectory, ${\mathcal O}(|x-\psi(t)|^{-2})$. Therefore 
$F={\mathcal O}(|x-\psi(t)|^{-1})$, the derivatives $f_{,a}$ are 
${\mathcal O}(|x-\psi(t)|^{-2})$
Thus $g_{,a}$  should be ${\mathcal O}(|x-\psi(t)|)$ - a regular term.\\
Since $f$ is singular the construction is not straightforward.
\end{document}